\documentclass[aps,prl,twocolumn, reprint, amsmath, amssymb]{revtex4-1}
\usepackage[centering,hmargin=2cm,vmargin=2cm]{geometry}

\newcommand{\td}{\text{d}}
\def\be{\begin{equation}}
\def\ee{\end{equation}}
\def\bea{\begin{eqnarray}}
\def\eea{\end{eqnarray}}
\def\nn{\nonumber}

\begin{document}

\title{New thermodynamic identities for five-dimensional black holes}

\author{Hari K Kunduri}
\email{hkkunduri@mun.ca}
\affiliation{Department of Mathematics and Statistics, Memorial University of Newfoundland, St John's, Canada}

\author{James Lucietti}
\email{j.lucietti@ed.ac.uk}
\affiliation{School of Mathematics and Maxwell Institute of Mathematical Sciences, University of Edinburgh, King's Buildings, Edinburgh, UK}

\begin{abstract}
We derive new identities for the thermodynamic variables of five-dimensional, asymptotically flat, stationary and biaxisymmetric vacuum black holes.   These identities depend on the topology of the solution  and include contributions arising from certain topological charges. 
The proof employs the harmonic map formulation of the vacuum Einstein equations for solutions with these symmetries.
\end{abstract}

\maketitle
A fundamental result in the theory of equilibrium black holes are the laws of black hole mechanics~\cite{Bardeen:1973gs}. These are formally analogous to the laws of thermodynamics, upon identifying the mass $M$, area $A_H$, surface gravity $\kappa$ of the black hole, with the energy, entropy and temperature, respectively. Famously, Hawking showed that quantum mechanically this is in fact a physical equivalence: black holes radiate at a small temperature and possess a large entropy \cite{Hawking:1974sw}. This profound result has since guided studies of quantum gravity.

The study of higher dimensional General Relativity has received much attention, due to its emergence in modern approaches to quantum gravity~\cite{Emparan:2008eg}. Interestingly, although the black hole uniqueness theorem no longer holds~\cite{Emparan:2001wn}, the laws of black hole mechanics remain valid~\cite{Myers:1986un}. In particular, the first law is closely related to the Smarr relation~\cite{Smarr:1972kt}, which for five-dimensional asymptotically flat, stationary and biaxisymmetric vacuum black holes, is
\be\label{Smarr}
M = \frac{3 \kappa A_H}{16 \pi} + \frac{3}{2} \Omega_i J_i  \; ,
\ee
where $\Omega_i, J_i$, for $i=1,2$, are the angular velocities and momenta relative to orthogonal planes at infinity.

In fact, the classification of black hole solutions in this class is an open problem. A uniqueness theorem has been established which reveals that the extra data required to specify such black holes is given by the so called rod structure~\cite{Harmark:2004rm, Hollands:2007aj, Hollands:2008fm}. This is certain data defined on the orbit space $\hat{M} = \mathcal{M}/ (\mathbb{R}\times U(1)^2)$ and in particular encodes both the horizon and spacetime topology. However, the issue of which rod structures lead to regular black hole solutions of the Einstein equations is not understood. In particular, the existence of a vacuum black hole with lens space topology (a black lens) or a domain of outer communication (DOC) with nontrivial topology remain unclear (see~\cite{Khuri:2017xsc} for recent progress).

In fact, asymptotically flat, stationary and biaxisymmetric, black hole solutions with such topology are known to exist in supergravity (Einstein-Maxwell theory coupled to a Chern-Simons term)~\cite{Kunduri:2014kja, Kunduri:2014iga, Horowitz:2017fyg}. These are supersymmetric and recently a full classification of such solutions has been derived, revealing a rich moduli space of spherical black holes, black rings and black lenses in spacetimes with noncontractible 2-cycles~\cite{Breunholder:2017ubu}.   Although nonsupersymmetric examples of such solutions are lacking, it has been shown that the first law of black hole mechanics (and the Smarr relation) include magnetic flux terms which are defined on any 2-cycles~\cite{Kunduri:2013vka}.  Thus in hindsight, it may seem surprising that for vacuum black holes, the first law and Smarr relations take a universal form independent of the topology. 

The purpose of this note is to present new identities for the thermodynamic variables of asymptotically flat, stationary and biaxisymmetric vacuum black holes. Furthermore, these identities do depend on the topology and include contributions from certain purely gravitational topological charges.  They arise as a consequence of the $SL(3,\mathbb{R})$ hidden symmetry of the vacuum Einstein equations for solutions in this class. Even for the explicitly known solutions, our identity is nontrivial, and to the best of our knowledge new. 

 For the  Myers-Perry black hole~\cite{Myers:1986un} we find
\begin{equation}\label{MPiden}
 \Omega_1 J_2 + \Omega_2 J_1-  \frac{16 M  \Omega_1 \Omega_2}{9\pi} \left(  M + \frac{3\kappa A_H}{16\pi}\right)=0   \; ,
\end{equation}  
whereas for the black ring~\cite{Pomeransky:2006bd}
\begin{align}\label{BRiden}
\Omega_1 J_2 +  \Omega_2 J_1 &- \frac{16 M \Omega_1 \Omega_2}{9\pi} \left(M + \frac{3\kappa A_H}{16\pi}\right)  \\ &=
 \frac{2}{3}M \Omega_1 \mathcal{G}_\xi[D] + \frac{ \kappa A_H}{8\pi} \Omega_1 \mathcal{G}_k[D] \nonumber  \; ,
 \end{align}  
 where $\Omega_1$ is the angular velocity along the $S^1$ of the ring and we have introduced the fluxes
\begin{equation}\label{flux}
\mathcal{G}_\zeta[D] = \frac{1}{2\pi} \int_{[D]} \star (k \wedge \td \zeta) \;  ,
\end{equation}
defined on the noncontractible 2-disc $D$ which ends on the horizon and $k, \xi$ are the stationary and corotating Killing fields respectively. These fluxes may be evaluated on any 2-surface homologous to $D$ since the integrand is closed by the Einstein equations and thus define topological `charges'. We have also verified these identities using the explicit form of the solutions. These are nontrivial only for {\it doubly} spinning black holes (i.e. $\Omega_i\neq 0$ for $i=1,2$). Indeed, for the black ring, the singly spinning limit is $\Omega_2\to 0$ and implies $\mathcal{G}_\zeta[D]\to 0$; this is because for the black ring $\mathcal{G}_\zeta[D] = \int_I \star (k \wedge m_1 \wedge \td \zeta)$, where $I \cong D/U(1)$ is the orbit space of $D$ under the $U(1)$ generated by $m_1$, and in the limit $\Omega_2\to 0$ the isometry group generated by the span of $k,m_1$ is orthogonally transitive.  In general, our new identities constrain generic doubly spinning black holes. 
 
 We now give a sketch of the proof of the general identities. Consider a five-dimensional asymptotically flat black hole spacetime $(\mathcal{M}, \mathbf{g})$ with an isometry group $\mathbb{R} \times U(1)^2$ generated by the stationary Killing vector field $k$ and two rotational Killing fields $m_i$, $i=1,2$ (spacelike with $2\pi$ periodic orbits). The event horizon must be a Killing horizon with respect to the corotating Killing field $\xi = k + \Omega_i m_i$~\cite{Figueras:2009ci}. We will assume the horizon is nondegenerate. The axis is the set of points for which $\det \gamma_{ij}=0$ where $\gamma_{ij} = \mathbf{g}(m_i, m_j)$.  It has been shown that the orbit space $\hat{M}$ is a simply connected 2d manifold with boundaries and corners~\cite{Hollands:2007aj}. The function $\rho= \sqrt{-\det G}$ where $G$ is the $3 \times 3$ matrix of inner products of $k,m_i$ is harmonic on $\hat{M}$. Furthermore, it has been shown that $\rho>0$ in the interior of $\hat{M}$ and $\rho=0$ only on the axis or horizon~\cite{Chrusciel:2008rh}.  This allows one to use $\rho$ and its harmonic conjugate $z$, defined by $\td z = - \star_2 \td \rho$, as global coordinates on $\hat{M}$ and hence identify the orbit space with the half-plane $\{ (\rho, z)\, | \, \rho>0 \}$ . 
 
The boundary segments of $\hat{M}$ are intervals on the $z$-axis that correspond to either a horizon or an axis on which $\gamma_{ij}$ is rank-1, whereas the corners of $\hat{M}$ are at the endpoints of adjacent axis intervals where $\gamma_{ij}$ is rank-0. We assume a connected horizon, in which case the $z$-axis divides into a horizon rod $H=[z_0,z_1]$, axis rods $I_A=[z_A,z_{A+1}]$ with $A\neq 0$, $-m\leq A \leq n-1$, ordered so $ z_{A}<z_{A+1}$, and semi-infinite axis rods $I_+ \equiv I_n=[z_n, \infty) $ and $I_- \equiv I_{-m-1}=(-\infty, z_{-m}]$. $z_{A\neq 0,1}$ are the corners. The rotational Killing fields $m_i$ are defined up to an $SL(2,\mathbb{Z})$ transformation. We choose these so that $m_1$ and $m_2$ vanish on $I_+$ and $I_-$ respectively. On every axis rod $I_A$ a vector $v_A=v_A^im_i$ vanishes (i.e. $v_A^i \in \text{null}(\gamma_{ij})$), where $v^i_A$ are coprime integers. We denote the union of all axis rods, including $I_\pm$, by $I$.

 It is well known that the vacuum Einstein equations for 5d spacetimes admitting {\it two} commuting Killing fields $\zeta_\mu$, $\mu=0,1$, are equivalent to a 3-dimensional theory of gravity coupled to a harmonic map whose target space is an $SL(3,\mathbb{R})$-coset, provided the $2\times 2$ matrix $\beta_{\mu \nu} = \mathbf{g}(\zeta_\mu, \zeta_\nu)$ is invertible~\cite{Maison:1979kx} (this generalises the harmonic map equations for 4d vacuum spacetimes with a Killing field, see e.g.~\cite{Geroch:1970nt}).  
For black hole spacetimes with three commuting Killing fields as above, there could be many ways of reducing to such a 3d theory: we find there are two natural choices. The obvious choice is to reduce on the $m_i$ ($\gamma_{ij}$ is positive definite away from the axes), which is in fact what is used to prove the uniqueness theorems~\cite{Hollands:2007aj, Figueras:2009ci}. 

A less obvious choice, which will be key to establishing the new identities, is to choose $\zeta_{\mu}$ to be the stationary and corotating Killing fields $k, \xi$.  Indeed, it is not even obvious that in this case $\beta_{\mu\nu}$ is invertible. Clearly, this can only be the case if the black hole is rotating (which we assume henceforth).    

{\it Conjecture:} For any rotating black hole $\det \beta_{\mu \nu}<0$ in the DOC away from the axis, where $\zeta_\mu$ are the stationary and corotating Killing field.

We now give evidence in favour of this conjecture. Firstly, one can write $\beta\equiv \det \beta_{\mu \nu}= |k|^2 |\xi|^2 - (k \cdot \xi)^2$ as
\begin{equation}\label{beta}
\beta =  q_{ij} \Omega_i \Omega_j, \qquad q_{ij} \equiv  g_{00} g_{ij} - g_{0i} g_{0j}.
\end{equation} 
Furthemore, the identity $\det q_{ij}  = - g_{00} \rho^2$ holds.  Thus outside an ergoregion $g_{00} < 0$ (and away from the axis), $q_{ij}$ is negative definite so we must have $\beta<0$. The ergosurface $g_{00}=0$ is timelike so it must be that on this $g_{0i} \neq 0$ and again \eqref{beta} implies $\beta < 0$. On the horizon $\xi$ is null and $k$ is spacelike and tangent to the horizon so $\beta=0$. It follows that just outside a nondegenerate horizon $| \xi |^2<0$ and $| k |^2>0$ so that again $\beta<0$ (using~\cite{Kunduri:2008rs} we have also checked that near a degenerate horizon $\beta<0$ even though $\xi$ is not always timelike). Thus to prove the above conjecture one needs to show that $\beta<0$ in the rest of the ergoregion. 

Now consider the axis. At points corresponding to the corners of $\hat{M}$ we have $m_1=m_2=0$ and hence $\beta=0$. Next, consider an axis rod $I_A$ and let $v_A= (p,q)$ be the corresponding vanishing Killing field written in the $(m_1,m_2)$ basis. Then, $w_A=( r, s)$, where $ps-qr=1$, is nonvanishing on $I_A$ and
\begin{equation} 
q_{ij} \Omega_i \Omega_j|_{I_A} = (|k|^2 |w_A|^2 - (k \cdot w_A)^2)\left(q \Omega_1 -  p \Omega_2\right)^2 \; .  \label{qaxis}
\end{equation}  
 Now, the span of $k,w_A$ must be timelike on $I_A$ (since it is on the axis and in the DOC~\cite{Chrusciel:2008rh}), so the first factor in (\ref{qaxis}) is negative and hence \eqref{beta} implies $\beta <0$ provided $\Omega_1,\Omega_2$ are both nonvanishing and incommensurate. We deduce that for such {\it generic} rotating black holes $\beta<0$ on the axis except at points corresponding to the corners of the orbit space. This lends further support to our conjecture (i.e. by continuity we deduce $\beta<0$ everywhere near the axis).  Henceforth, we will consider generic rotating black holes and assume the validity of our conjecture.

We now introduce coordinates $(u,v,w)$ adapted to the three commuting Killing fieds $k, \xi, m$, where $m= c_i m_i$ for some arbitrary constants $c_i$ such that $\alpha \equiv c_1 \Omega_2 - c_2 \Omega_1 > 0$ (to ensure their linear independence), so $k = \partial_u, \xi = \partial_v , m= \partial_w$. Furthermore, since the distribution orthogonal to the span of $k, \xi, m$ must be integrable, these coordinates can be chosen so that 
\begin{equation}
\label{KKmetric}
\mathbf{g} = \beta_{\mu \nu} (\td x^\mu + B^\mu \td w)(\td x^\nu + B^\nu \td w) +  \frac{\tilde \rho^2 \td w^2}{|\beta|}  + g_2 
\end{equation} 
where $(x^\mu)=(u,v)$, $\tilde \rho = \alpha \rho$ and $g_2\propto \td \rho^2+ \td z^2$ is the metric on $\hat{M}$. The $SL(3,\mathbb{R})$-coset harmonic map equations arising from the reduction along $k, \xi$~\cite{Maison:1979kx}, when further reduced on the third Killing field $m$ to the 2d orbit space $\hat{M}$, are  $\td \mathcal{J}=0$ where $\mathcal{J}=\tilde{\rho} \Phi^{-1} \star_2 \td \Phi$ and
\begin{equation} 
\Phi = 
\begin{pmatrix} \beta_{\mu \nu} + \beta^{-1}U_\mu U_\nu & - \beta^{-1}U_\mu  \\ -\beta^{-1} U_\nu & \beta^{-1} \end{pmatrix}
\end{equation} and $U_\mu $ are smooth potentials defined by
\begin{equation}\label{dU}
\td U_\mu \equiv \star (k \wedge \xi \wedge \td \zeta_\mu)  \; .
\end{equation} 
Since $\text{Tr} \mathcal{J}=0$, the harmonic map equations are equivalent to $\td \mathcal{J}^\mu_{~\nu}= \td \mathcal{J}_\mu=0$ where
\begin{align} 
\mathcal{J}^\mu_{~\nu} &\equiv \tilde \rho \beta^{\mu \alpha} \star_2 \td \beta_{\alpha \nu} + B^\mu \td U_\nu  + \td C^\mu_{~\nu} \label{tensor}\\
\mathcal{J}_\nu &\equiv \tilde \rho U_\beta \beta^{\beta\alpha} \star_2 \td \beta_{\alpha \nu} - U_\nu U_\beta \td B^\beta \nonumber \\ & \quad + \tilde \rho U_\nu \beta^{-1}\star_2 \td \beta - \tilde \rho \star_2 \td U_\nu + \td C_\nu \label{vector}
\end{align} 
and 
\begin{equation} 
\label{dB} \td B^\mu = -\beta^{-1} \tilde\rho \beta^{\mu\nu} \star_2 \td U_\nu 
\end{equation}
(the latter follows from the definition of $U_\mu$), and $C^\mu_{~\nu}, C_\mu$ are arbitrary functions introduced for later convenience.  Our new identities are obtained by integrating the one-forms $\mathcal{J}^\mu_{~\nu}, \mathcal{J}_{\nu}$ over $\partial \hat{M}$ which by Stokes' theorem must vanish.  

Asymptotic flatness of $(\mathcal{M},\mathbf{g})$ fixes a particular asymptotic expansion for the metric components  \eqref{KKmetric}. To express this it is convenient to choose coordinates $(t, \phi_i, R, \theta)$ such that $k= \partial_t, m_i = \partial_{\phi_i}$, $\rho = \tfrac{1}{2} R^2 \sin (2 \theta)$,  $z-z_*= \tfrac{1}{2} R^2 \cos( 2\theta)$ where $R>0$,  $\theta \in [0,\pi/2]$ and $z_*$ is an arbitrary constant. The metric has an asymptotic expansion in powers of $R^{-2}$ for $R \to \infty$ given by~\cite{Harmark:2004rm, Hollands:2008fm}
\begin{align}\label{AFg}
\mathbf{g} \sim  &-\left( 1- \frac{8M}{3\pi R^2}  \right)\td t^2 - \sum_{i=1}^2\frac{8}{\pi R^2}J_i \mu^2_i \td t \td \phi_i \nn \\ \nn
&+ \sum_{i=1}^2 \left(1 + \frac{4(M-(-1)^i \eta)}{3\pi R^2}\right)R^2 \mu_i^2 \td \phi_i^2 \\
&+ \frac{16 \zeta \sin^2\theta \cos^2\theta}{R^2}  \td \phi_1 \td\phi_2+ \td R^2 +R^2 \td{\theta}^2
\end{align}
where $\mu_i= (\sin\theta, \cos\theta)$ and $M, J_i$ are the mass and angular momenta and $\eta, \zeta$ are constants.   The constant $\eta$ depends on the choice of integration constant $z_*$ in the harmonic conjugate $z$ to $\rho$ ($\zeta$ is gauge invariant). We will fix this gauge so that $z_*=z_0$ where $[z_0,z_1]$ is the horizon rod. Below we will find the corresponding value of $\eta$.   
 The  asymptotic expansions of $\beta_{\mu\nu}, B^\mu$ in \eqref{KKmetric} can be deduced from \eqref{AFg}. The orbit space $(\hat{M}, g_2)$ inherits an asymptotic end as $R\to \infty$, where $g_2 \sim \td R^2+ R^2 \td \theta^2$,  with boundary  $S_\infty$ defined by $R=$ constant $\to \infty$ corresponding to the `semi-circle at infinity'.

We now evaluate the integral of \eqref{tensor} over $S_\infty$.
The asymptotics \eqref{AFg} imply $\tilde\rho \beta^{\mu \alpha} \star_2 \td \beta_{\alpha \nu}=  [O(R^2) \td \theta +O(R) \td R ]s^\mu_{~\nu}$, where $s^\mu_{~\nu} = (\xi^\mu- k^\mu) \xi_\nu^*$ and $\xi^*_\mu, k^*_\mu$ are the dual vectors to $\xi^\mu, k^\mu$. Similarly, inverting \eqref{dB} gives
\begin{align}
U_u  &\sim   \frac{4}{\pi R^2}(J_1 \Omega_2 \cos^2\theta+ J_2 \Omega_1 \sin^2\theta) \\
U_v  &\sim  \Omega_1 \Omega_2 R^2 - \frac{4\eta \Omega_1 \Omega_2 \cos2\theta}{3\pi} 
\end{align} 
where we have fixed the integration constants. This implies $B^\mu \td U_\nu = [O(1) \td \theta+ O(R)\td R ]s^\mu_{~\nu}$. The resulting divergent terms in $\mathcal{J}^\mu_{~\nu}$ can be removed by the counterterm
\begin{gather}
C^\mu_{~\nu} = \left(\alpha (z-z_0) - \tilde{\alpha} \sqrt{\rho^2 + (z-z_0)^2} \right)  s^\mu_{~\nu}  \; ,    \label{counter}
\end{gather} 
where $\tilde{\alpha}=c_1 \Omega_2+c_2\Omega_1$, which we assume henceforth.
The integral of $\mathcal{J}^\mu_{~\nu}$ over the semi-circle at infinity now converges and evaluates to
\begin{equation}
\int_{S_\infty} \mathcal{J}^\mu_{~\nu} =\begin{pmatrix} \frac{8M\alpha}{3\pi} & \frac{4\alpha}{\pi}(M - \Omega_i J_i  + c \eta) \\ 0 & -\frac{4\alpha}{3\pi}(M + 3 c \eta) \end{pmatrix}
\end{equation} where $c = -\tilde{\alpha}/(3 \alpha)$. 

Next we consider the integral of \eqref{tensor} over the horizon rod $H$. Recall on the horizon $\rho = 0$ and the matrix $\beta_{\mu \nu}$ is noninvertible, so one must take care when evaluating the integrand.  Near a rotating nondegenerate horizon, $z \in H$ and $\rho \to 0$, we can write (see e.g.~\cite{Harmark:2004rm, Hollands:2008fm})
\begin{equation}
\beta_{\mu \nu} = a \rho^2 \xi^*_\mu \xi^*_\nu + 2b\rho^2 \xi^*_{(\mu}k^*_{\nu)} + c k_\mu^* k^*_\nu,
\end{equation} for some smooth functions $a, b, c$ satisfying $ ac - b^2 \rho^2 < 0$. Using this we find
\begin{equation}
\tilde\rho \beta^{\mu \alpha}\star_2 \td \beta_{\alpha \nu}|_{H}= -2\alpha \frac{\xi^\mu\xi^\alpha \beta_{\alpha \nu} }{|\xi|^2} \td z
\end{equation}  
where we have made use of the completeness relation $\delta^\mu_{~\nu} = \xi^\mu \xi^*_\nu + k^\mu k^*_\nu$.
Similarly, near the horizon we can write $g_{\mu w} = e \rho^2 \xi^*_{\mu}+ f k^*_\mu$ for functions $e,f$. This implies that $B^\mu= \beta^{\mu\nu} g_{\nu w}$ is in fact regular on the horizon, so \eqref{dB} shows that $U^H_\mu\equiv U_\mu|_H$ are constant on the horizon.  In total we find,
\begin{equation}
\int_{H} \mathcal{J}^\mu_{~\nu} = -2\alpha \xi^\mu\int_{H} \frac{\xi^\alpha\beta_{\alpha \nu}}{|\xi|^2} \td z - 2c_2 \Omega_1 \ell_H  s^\mu_{~\nu} 
\end{equation}  
where $\ell_H = z_1-z_0$. 

Evaluating the integral over the axis is more subtle. As explained above, for a generic doubly rotating black hole (i.e. $\Omega_1, \Omega_2$ incommensurate), then $\beta<0$ everywhere on the axis except at the corners of $\hat{M}$. In fact, since the span of $k,m_i$ is timelike in the DOC~\cite{Chrusciel:2008rh}, at any corner we see that $\beta_{\mu\nu}$ is rank-1 (i.e. $k\neq 0$). To analyse the behaviour of the integrand (\ref{tensor}) on the axis we need the geometry of spacetime near a corner of the orbit space. It does not appear this has been discussed before for vacuum gravity (for supergravity see~\cite{Breunholder:2017ubu}).

Let $z=z_A$ be a corner of $\hat{M}$. The rod vectors which vanish on the adjacent axis rods $I_{A-1}$ and $I_A$  must obey the compatibility condition $\det( v_{A-1},  v_A) = \pm 1$ \cite{Hollands:2007aj}. This implies that a neighbourhood of a corner is diffeomorphic to $\mathbb{R}^{1,4}$. Using normal coordinates at $z_A$, one can then introduce spherical coordinates such that near the corner $r\to 0$,  $\vartheta \in [0,\pi/2]$, and
\begin{equation}
\mathbf{g} \sim -a^2 \td t^2 + \td r^2 + r^2\left(\td \vartheta^2 + \sin^2\vartheta \td \phi_R^2 + \cos^2\vartheta \td \phi_L^2\right) \nn
\end{equation} where $a$ is a constant, $k = \partial_t$, $v_{A-1}=\partial_{\phi_L}, v_A=\partial_{\phi_R}$.  It follows that near a corner $\tilde\rho \beta^{\mu \alpha} \star_2 \td \beta_{\alpha \nu} = [ O(r^2)\td \vartheta+O(r) \td r] s^\mu_{~\nu}$.  In the interior of any axis rod  clearly $\tilde\rho \beta^{\mu \alpha} \star_2 \td \beta_{\alpha \nu} = 0$ (since $\rho=0, \beta<0$), so we deduce $\int_{I} \tilde\rho \beta^{\mu \alpha} \star_2 \td \beta_{\alpha \nu} =0$. Also, the functions $U_\mu$ are smooth and hence \eqref{dB} implies $B^\mu_A\equiv B^\mu|_{I_A}$ is constant in the interior of any axis rod $I_A$.  However, working in the above chart,  we find that $B^\mu$ on $I$ jumps across each corner $z=z_A$,
\begin{align}\label{Bjump}
&B^\mu_{A-1} - B^\mu_{A} = \alpha b_A (\xi^\mu- k^\mu) \; , \\ 
&b_A  = \pm \left[(\Omega_1 v^2_{A-1} - \Omega_2 v^1_{A-1})(\Omega_1 v_A^2 - \Omega_2 v_A^1)\right]^{-1}. \nn
\end{align} 
We find the axis integral
\begin{align}
\int_I \mathcal{J}^\mu_{~\nu} &= (B^\mu|_{z_0}- B^\mu|_{z_1} ) U_\nu^H  - \frac{4 \eta \Omega_1 \Omega_2 \xi^*_\nu}{3\pi} (B_+^\mu+B_-^\mu)\nn \\
&+ \sum_{A\neq 0,1} (B^\mu_{A-1} - B^\mu_{A} ) U_\nu|_{z_A} + 2 c_2 \Omega_1 \ell_H  s^\mu_{~\nu}
\end{align}
where the sum over $A\neq 0,1$ is equivalent to the sum over corners, and $B_{\pm}^\mu = B^\mu|_{I_{\pm}}$. From the asymptotics $B_+^\mu =  (\xi^\mu- k^\mu)c_2/\Omega_2$ and $B_-^\mu =  (\xi^\mu- k^\mu)c_1/\Omega_1$.

We may now deduce our first set of identities.  As mentioned above, by Stokes' theorem the sum of the integrals of $\mathcal{J}^\mu_{~\nu}$ over the axis, horizon and semi-circle at infinity must vanish. This yields a $2\times 2$ matrix of identities. The sum of the $uv$ and $vv$ components gives the Smarr relation \eqref{Smarr}. The $uu$ and $vv$ components give
\begin{gather} 
 b_H U_\mu^H  +  \sum_{A\neq 0,1} b_A U_\mu|_{z_A}    \label{iden1}
= \frac{4}{3\pi}\begin{pmatrix} 2M \\ M + \frac{3 \kappa A_H}{8\pi} \end{pmatrix}   \; ,
\end{gather} 
where $b_H= (\Omega_1\Omega_2)^{-1}- \sum_{A\neq 0,1} b_A$ is defined by $B^\mu|_{z_0} - B^\mu|_{z_1} = \alpha b_H (\xi^\mu-k^\mu)$,
and we used $\ell_H = \kappa A_H/(4\pi^2)$\cite{Hollands:2008fm}.
Finally, the $vu$ component gives a remarkably simple formula for the mass (upon use of  \eqref{iden1}),
\begin{equation}\label{Mass}
M = \frac{3\pi}{4}\int_H \lim_{\rho\to 0}  \frac{ k \cdot \xi}{|\xi|^2} \td z  \; .
\end{equation}
Curiously, we have verified that this mass formula also holds for the 4d Kerr solution (with a prefactor of $\tfrac{1}{2}$).
In summary, the identities arising from integrating $\td \mathcal{J}^\mu_{~\nu}=0$ over $\hat{M}$ are equivalent to (\ref{Smarr}), (\ref{iden1}) and \eqref{Mass}.

We now turn to evaluating the identities arising from integrating \eqref{vector} over $\partial \hat{M}$. We will focus on the $u-$component (the $v-$component only fixes subleading parts in the asymptotic expansion \eqref{AFg}).  We find that for convergence of the integral on the semi-circle at infinity no counterterm is needed ($C_\nu =0$) and
\begin{equation}
\int_{S_\infty} \mathcal{J}_u = \frac{4\alpha}{\pi}( \Omega_1 J_2 +  \Omega_2 J_1).
\end{equation} The horizon integral evaluates to (using \eqref{Mass})
\begin{align}\label{hor:vector}
\int_H \mathcal{J}_u &= -\alpha \left(\frac{8M}{3\pi}U_v^H+ \frac{\kappa A_H}{2\pi^2} U_u^H \right) \nonumber \\ &+ U_u^H U_\alpha^H (B^\alpha|_{z_0} - B^\alpha|_{z_1}).
\end{align} 
For the axis integral we find
\begin{equation}
\int_{I}\mathcal{J}_u  = \sum_{A\neq 0,1}\alpha b_A U_u|_{z_A} (-U_u|_{z_A} + U_v|_{z_A}) ,
\end{equation}
which may be evaluated by noting
\begin{equation}
\td B^\mu|_{I} =  -\alpha (\xi^\mu-k^\mu) \sum_{A\neq 0,1} b_A \delta(z-z_A) \td z  \; .
\end{equation} 
(To see this write $B^\mu|_I$ in terms of step functions).
The sum of these integrals again must vanish by Stokes' theorem (recall $\td \mathcal{J}_u=0$), and we obtain the identity
\begin{equation}
\begin{aligned}
 \Omega_1 J_2 &+  \Omega_2 J_1 = \frac{2M}{3}U_v^H + \frac{\kappa A_H U_u^H}{8\pi}  \\ &+ \frac{\pi}{4}\sum_{A\neq 0,1} b_A U_u|_{z_A}(U_u|_{z_A} - U_v|_{z_A}) \\ &+ \frac{\pi}{4}b_H  U_u^H (U_u^H - U_v^H)  \; .\label{iden2}
\end{aligned}
\end{equation} 
Our final identity may be now obtained by combining \eqref{iden1} and \eqref{iden2} as follows.

First, using $b_H$, write \eqref{iden1} as
\be
U^H_\mu = \frac{4 \Omega_1 \Omega_2}{3\pi} \left( \begin{array}{c} 2M \\  M+ \frac{3 \kappa A_H}{8\pi} \end{array} \right) -  \Omega_1 \Omega_2 \sum_{A\neq 0,1} b_A \Delta U^A_\mu
\ee
where $\Delta U^A_\mu = U_\mu|_{z_A}- U_\mu^H$.
Using this, we can then write \eqref{iden2} purely in terms of $\Delta U^A_\mu$ as,
\bea
&& \Omega_1J_2 +\Omega_2 J_1 = \frac{16 M \Omega_1 \Omega_2}{9\pi} \left( M+ \frac{3\kappa A_H}{16\pi} \right)   \label{mainid}\\
&&-  \sum_{A\neq 0,1} \left(\frac{\kappa A_H \Omega_1 \Omega_2}{8\pi} b_A \Delta U^A_u  - \frac{2M \Omega_1 \Omega_2}{3} b_A \Delta U^A_v  \right) \nn \\
&&+\frac{\pi}{4} \sum_{A, B \neq 0,1} b_A (\delta_{AB} -\Omega_1 \Omega_2  b_B ) \Delta U^A_u[ \Delta U^B_u- \Delta U_v^B] \nn   \; .
\eea
The differences $\Delta U^A_\mu $ can all be related to the topological charge \eqref{flux} using
\begin{equation}
 \int_{I_A} \td U_\mu  = (\Omega_2 v_A^1- \Omega_1 v_A^2)  \mathcal{G}_{\zeta_\mu}[C_A]   \label{fluxid}
\end{equation} 
where $C_A$ is the 2-cycle corresponding to the axis rod $I_A$ (with induced orientation) and we define $\mathcal{G}_{\zeta}[C]= \tfrac{1}{2\pi}\int_{[C]} \star ( k \wedge \td \zeta)$. Then, clearly $\Delta U_\mu^A= \sum_{B=1}^{A-1} \int_{I_B} \td U_\mu$ if $A>0$ (and similarly for $A<0$).  Thus the identity \eqref{mainid} together with \eqref{fluxid} relate the standard thermodynamic variables of a black hole to the rod structure and topology of the spacetime.  This is our main result.

For the Myers-Perry interval structure $I_- \cup H \cup I_+$ there are no corners, 
so \eqref{mainid} reduces to \eqref{MPiden}.  Now consider a more general interval structure $I_-\cup H \cup I_1 \cup I_+$ with rod vector $v_1=(p,1)$ and $p \in \mathbb{Z}$. For $p=0$ this corresponds to a black ring, whereas for $p\neq 0$ it corresponds to yet to be constructed black holes with $L(p,1)$ horizon topology (i.e. black lenses). In these cases there is one corner $z_2$, with $b_2=[(\Omega_1- p\Omega_2)\Omega_2]^{-1}$ and hence \eqref{mainid} and \eqref{fluxid} give
 \begin{align}
  \Omega_1 J_2 +  \Omega_2J_1 &- \frac{16 M \Omega_1 \Omega_2}{9\pi}\left(M + \frac{3\kappa A_H}{16\pi}\right)  \\
 &= \frac{2}{3} M \Omega_1 \mathcal{G}_\xi[D] + \frac{\kappa A_H}{8\pi} \Omega_1 \mathcal{G}_k[D] \nn \\ &\quad+\frac{\pi p}{4} \mathcal{G}_k[D](\mathcal{G}_\xi[D] - \mathcal{G}_k[D]) \nonumber  \; ,
 \end{align}
 where $D$ is the 2-disc corresponding to $I_1$. If $p=0$ this reduces to the black ring identity \eqref{BRiden}.  For $p \neq 0$ this gives a prediction for black lenses (if they exist). Analogous identities can be derived from \eqref{mainid}, \eqref{fluxid} for black holes with other topologies, e.g. $S^3$ black holes with 2-cycles in the DOC.   
 
It is interesting to consider analogous identities arising from the harmonic map defined by first reducing on the biaxial Killing fields $m_i$. The method is completely analogous to the above with $\beta_{\mu\nu}$ and $U_\mu$ replaced by $\gamma_{ij}$ and the twist potentials $Y_i$ (and $\det \gamma_{ij}>0$ away from the axis in the DOC). From the analogue of \eqref{tensor} we obtain four identities which are equivalent to (\ref{Smarr}),
 a formula for the gauge dependent constant
\begin{equation}
\begin{aligned}
\eta &= \frac{3\kappa A}{16\pi} + \frac{3}{2} (\Omega_1 J_1 - \Omega_2 J_2)  \\ &+ \frac{3\pi}{4} \sum_{I_A \neq I_\pm } \left( \text{sgn}(A) \ell_A +  \int_{I_A}\frac{\tilde{v}_A \cdot v_A}{|v_A|^2} \td z \right)
\end{aligned}
\end{equation} where $\ell_A = z_{A+1}-z_A$, $\tilde{v}_A^i = (-v_A^1, v_A^2)$, 
and
\begin{equation}
 \Omega_i J_j = \frac{\pi}{2} \sum_{I_A} v_A^i \int_{I_A} \lim_{\rho \to 0} \frac{m_j \cdot v_A}{|v_A|^2} \td z, \quad i \neq j . \label{iden3}
\end{equation}
The analogue of \eqref{vector} determines certain subleading constants (analogous to $\eta$) in the asymptotics of $Y_i$. In particular, \eqref{iden3} is a new identity relating the spin of the black hole to the rod structure. 

In summary, we have presented new identities relating the physical charges and spacetime topology of equilibrium, rotating black holes. In the context of black hole thermodynamics, our simplest identity \eqref{MPiden} gives a new nonlinear equation of state for the basic thermodynamic variables.
Our method clearly generalises to others theories of gravity which admit a coset harmonic map formulation. In particular, more general thermodynamic nonlinear identities should hold in the supergravity theories that govern the low-energy dynamics of string theory (for static Kaluza-Klein black holes in 11d supergravity see~\cite{Hollands:2012cc}). It would be interesting to understand the microscopic origin of such identities.

 {\bf Acknowledgements}. HKK is supported by NSERC Discovery Grant RGPIN-2018-04887.  JL is funded in part by STFC [ST/L000458/1].


\begin{thebibliography}{99}
  
\bibitem{Bardeen:1973gs}
  J.~M.~Bardeen, B.~Carter and S.~W.~Hawking,
  Commun.\ Math.\ Phys.\  {\bf 31} (1973) 161.  
  
\bibitem{Hawking:1974sw}
  S.~W.~Hawking,
  Commun.\ Math.\ Phys.\  {\bf 43} (1975) 199  
  
    \bibitem{Emparan:2008eg}
  R.~Emparan and H.~S.~Reall,
  Living Rev.\ Rel.\  {\bf 11} (2008) 6
  
  \bibitem{Emparan:2001wn}
  R.~Emparan and H.~S.~Reall,
  Phys.\ Rev.\ Lett.\  {\bf 88} (2002) 101101

 \bibitem{Myers:1986un}
  R.~C.~Myers and M.~J.~Perry,
  Annals Phys.\  {\bf 172} (1986) 304. 
  
 \bibitem{Smarr:1972kt}
  L.~Smarr,
  Phys.\ Rev.\ Lett.\  {\bf 30} (1973) 71
   Erratum: [Phys.\ Rev.\ Lett.\  {\bf 30} (1973) 521]. 
   
\bibitem{Harmark:2004rm}
  T.~Harmark,
  Phys.\ Rev.\ D {\bf 70} (2004) 124002
   
  \bibitem{Hollands:2007aj}
  S.~Hollands and S.~Yazadjiev,
  Commun.\ Math.\ Phys.\  {\bf 283} (2008) 749
  
\bibitem{Hollands:2008fm}
  S.~Hollands and S.~Yazadjiev,
  Commun.\ Math.\ Phys.\  {\bf 302} (2011) 631
  
\bibitem{Khuri:2017xsc}
  M.~Khuri, G.~Weinstein and S.~Yamada,
Comm. Partial Differential Equations 43 (2018), no. 8, 1205-1241. 
   
   \bibitem{Kunduri:2014kja}
  H.~K.~Kunduri and J.~Lucietti,
  Phys.\ Rev.\ Lett.\  {\bf 113} (2014) no.21,  211101
  
\bibitem{Hollands:2012cc}
  S.~Hollands,
  Class.\ Quant.\ Grav.\  {\bf 29} (2012) 205009
 
\bibitem{Kunduri:2014iga}
  H.~K.~Kunduri and J.~Lucietti,
  JHEP {\bf 1410} (2014) 082
  
\bibitem{Horowitz:2017fyg}
  G.~T.~Horowitz, H.~K.~Kunduri and J.~Lucietti,
  JHEP {\bf 1706} (2017) 048
   
\bibitem{Breunholder:2017ubu}
  V.~Breunh\"older and J.~Lucietti,
Commun.\ Math.\ Phys.\  {\bf 365} (2019) no.2,  471
  
   
  \bibitem{Kunduri:2013vka}
  H.~K.~Kunduri and J.~Lucietti,
  Class.\ Quant.\ Grav.\  {\bf 31} (2014) 032001
  

 
 \bibitem{Pomeransky:2006bd}
  A.~A.~Pomeransky and R.~A.~Sen'kov,
  hep-th/0612005.
  
\bibitem{Figueras:2009ci}
  P.~Figueras and J.~Lucietti,
  Class.\ Quant.\ Grav.\  {\bf 27} (2010) 095001
  
\bibitem{Maison:1979kx}
  D.~Maison,
  Gen.\ Rel.\ Grav.\  {\bf 10} (1979) 717.
  
\bibitem{Geroch:1970nt}
  R.~P.~Geroch,
  J.\ Math.\ Phys.\  {\bf 12} (1971) 918.
  
\bibitem{Chrusciel:2008rh}
  P.~T.~Chrusciel,
  J.\ Math.\ Phys.\  {\bf 50} (2009) 052501
  
\bibitem{Kunduri:2008rs}
  H.~K.~Kunduri and J.~Lucietti,
  J.\ Math.\ Phys.\  {\bf 50} (2009) 082502
  


\end{thebibliography}
\end{document}